\documentclass[letter,longauth]{aa}  

\usepackage{graphicx}
\usepackage{txfonts}
\begin{document}

   \title{The ALMA-ALPINE [CII] survey: Kennicutt-Schmidt relation in four massive main-sequence galaxies at z$\sim$4.5}
   
    \authorrunning{B\'ethermin et al.}
    \titlerunning{KS relation at z$\sim$4.5}

   \author{M. B\'ethermin\inst{1,2} \and
          C. Accard\inst{1}\thanks{The second and third authors are master's students, who had a similar contributions to this letter (analysis of the  first three sources and initial results).} \and
          C. Guillaume\inst{1} \and  
          M. Dessauges-Zavadsky\inst{3} \and
          E. Ibar\inst{4} \and
          P. Cassata\inst{5,6} \and
          T. Devereaux\inst{5,6} \and
          A. Faisst\inst{7} \and
          J. Freundlich\inst{1} \and
          G.~C. Jones\inst{8} \and
          K. Kraljic\inst{1} \and
          H. Algera\inst{9,10} \and
          R.~O. Amor\'{i}n\inst{11,12} \and
          S. Bardelli\inst{19}
          M. Boquien\inst{13} \and
          V. Buat\inst{2} \and
          E. Donghia\inst{14} \and
          Y. Dubois\inst{15} \and
          A. Ferrara\inst{16} \and
          Y. Fudamoto\inst{29} \and
          M. Ginolfi\inst{30,31} \and
          P. Guillard\inst{15} \and
          M. Giavalisco\inst{17} \and
          C. Gruppioni\inst{19} \and
          G. Gururajan\inst{18,19}\and
          N. Hathi\inst{20} \and
          C.~C. Hayward\inst{21} \and
          A.~M. Koekemoer\inst{20} \and
          B.~C. Lemaux\inst{22,33}\and
          G.~E. Magdis\inst{23,24,25} \and
          J. Molina\inst{26} \and
          D. Narayanan\inst{23,27} \and
          L. Mayer\inst{32} \and
          F. Pozzi\inst{19} \and
          F. Rizzo\inst{23,25} \and
          M. Romano\inst{6,28} \and
          L. Tasca\inst{2} \and
          P. Theul\'e\inst{2} \and
          D. Vergani\inst{19} \and
          L. Vallini\inst{19} \and
          G. Zamorani\inst{19} \and 
          A. Zanella\inst{6} \and
          E. Zucca\inst{19}
           }
          
    \institute{Universit\'e de Strasbourg, CNRS, Observatoire astronomique de Strasbourg, UMR 7550, 67000 Strasbourg, France\email{matthieu.bethermin@astro.unistra.fr} \and
    Aix Marseille Univ, CNRS, CNES, LAM, Marseille, France \and
    Department of Astronomy, University of Geneva, Chemin Pegasi 51, 1290 Versoix, Switzerland \and
    Instituto de F\'isica y Astronom\'ia, Universidad de Valpara\'iso, Avda. Gran Breta\~na 1111, Valpara\'iso, Chile \and
    Dipartimento di Fisica e Astronomia, Università di Padova, Vicolo dell'Osservatorio 3, I-35122, Padova, Italy \and
    INAF – Osservatorio Astronomico di Padova, Vicolo dell’Osservatorio 5, I-35122, Padova, Italy \and
    IPAC, California Institute of Technology 1200 E California Boulevard, Pasadena, CA 91125, USA \and
    Department of Physics, University of Oxford, Denys Wilkinson Building, Keble Road, Oxford OX1 3RH, UK \and
    Hiroshima Astrophysical Science Center, Hiroshima University, 1-3-1 Kagamiyama, Higashi-Hiroshima, Hiroshima 739-8526, Japan \and
    National Astronomical Observatory of Japan, 2-21-1, Osawa, Mitaka, Tokyo, Japan \and
    ARAID Foundation. Centro de Estudios de F\'{\i}sica del Cosmos de Arag\'{o}n (CEFCA), Unidad Asociada al CSIC, Plaza San Juan 1, E--44001 Teruel, Spain \and
    Departamento de Astronom\'{i}a, Universidad de La Serena, Av. Juan Cisternas 1200 Norte, La Serena 1720236, Chile \and
    Instituto de Alta Investigación, Universidad de Tarapacá, Casilla 7D, Arica, Chile \and
    Astronomy Department, University of Wisconsin, Madison \and
    Institut d’Astrophysique de Paris, CNRS and Sorbonne Université, UMR 7095, 98 bis Boulevard Arago, F-75014 Paris, France \and
    Scuola Normale Superiore, Piazza dei Cavalieri 7, 50126 Pisa, Italy \and
    Astronomy Department, University of Massachusetts, Amherst, MA 01003, USA \and 
    University of Bologna - Department of Physics and Astronomy “Augusto Righi” (DIFA), Via Gobetti 93/2, I-40129, Bologna, Italy \and
    INAF - Osservatorio di Astrofisica e Scienza dello Spazio, Via Gobetti 93/3, I-40129, Bologna, Italy \and
    Space Telescope Science Institute, Baltimore, MD 21218, USA \and
    Center for Computational Astrophysics, Flatiron Institute, 162 Fifth Avenue, New York, NY 10010, USA \and
    Department of Physics and Astronomy, University of California Davis, One Shields Avenue, Davis, CA 95616, USA \and
    Cosmic Dawn Center (DAWN), Jagtvej 128, DK2200 Copenhagen N, Denmark \and
    DTU-Space, Technical University of Denmark, Elektrovej 327, DK2800 Kgs. Lyngby, Denmark \and
    Niels Bohr Institute, University of Copenhagen, Jagtvej 128, DK-2200 Copenhagen N, Denmark \and
    Department of Space, Earth and Environment, Chalmers University of Technology, Onsala Space Observatory, 439 92 Onsala, Sweden \and
    Department of Astronomy, University of Florida, 211 Bryant Space Sciences Center, Gainesville, FL 32611 USA \and
    National Centre for Nuclear Research, ul. Pasteura 7, 02-093, Warsaw, Poland \and
    Center for Frontier Science, Chiba University, 1-33 Yayoi-cho, Inage-ku, Chiba 263-8522, Japan \and
    Dipartimento di Fisica e Astronomia, Università degli Studi di Firenze, Via G. Sansone 1,I-50019, Sesto Fiorentino, Firenze, Italy \and
    INAF - Osservatorio Astrofisico di Arcetri, Largo E. Fermi 5, I-50125, Firenze, Italy \and
    Center for Theoretical Astrophysics and Cosmology, Institute for Computational Science, University of Zurich, Winterthurerstrasse 190, Zurich, Switzerland \and
    Gemini Observatory, NSF's NOIRLab, 670 N. A'ohoku Place, Hilo, Hawai'i, 96720, USA
    }

    \date{Received 29/09/2023; accepted 13/11/2023}

 
  \abstract{}
   {The Kennicutt-Schmidt (KS) relation between the gas and the star formation rate (SFR) surface density ($\Sigma_{\rm gas} - \Sigma_{\rm SFR}$) is essential to understand star formation processes in galaxies. To date, it has been measured up to z$\sim$2.5 in main-sequence galaxies. In this letter our aim is to put constraints at z$\sim$4.5 using a sample of four massive main-sequence galaxies observed by ALMA at high resolution.}
   {We obtained $\sim$0.3"-resolution [CII] and continuum maps of our objects, which we then converted into gas and obscured SFR surface density maps. In addition, we produced unobscured SFR surface density maps by convolving \textit{Hubble} ancillary data in the rest-frame UV. We then derived the average $\Sigma_{\rm SFR}$ in various $\Sigma_{\rm gas}$ bins, and estimated the uncertainties using a Monte Carlo sampling.}
   {Our galaxy sample follows the KS relation measured in main-sequence galaxies at lower redshift, and is slightly lower than the predictions from simulations. Our data points probe the high end both in terms of $\Sigma_{\rm gas}$ and $\Sigma_{\rm SFR}$, and gas depletion timescales (285-843\,Myr) remain similar to z$\sim$2 objects. However, three of our objects are clearly morphologically disturbed, and we could have expected shorter gas depletion timescales ($\lesssim$100\,Myr) similar to merger-driven starbursts at lower redshifts. This suggests that the mechanisms triggering starbursts at high redshift may be different than in the low- and intermediate-z Universe.}
   {}
%

\maketitle

\section{Introduction}

The Kennicutt-Schmidt (KS) empirical relation \citep{Schmidt1963,Kennicutt1998,De_los_Reyes2019} linking the gas and the star formation rate (SFR) surface densities ($\Sigma_{\rm gas} - \Sigma_{\rm SFR}$) is a key tool for understanding the star formation in galaxies across cosmic times. This relation has been vastly explored in the local Universe including at sub-galactic scales \citep[e.g.,][]{Leroy2013,Pessa2021,Sun2023}. These studies have also shown that  star formation is mainly correlated with the molecular gas \citep[e.g.,][]{Bigiel2008}.

 At higher redshift, these measurements are difficult, since they require     resolved maps of the molecular gas, usually using the CO rotational lines, together with reliable star formation maps \citep[e.g.,][]{Molina2019}. Since current observing facilities are not sensitive enough to detect the atomic hydrogen at 21\,cm in the high-redshift Universe, previous and current studies mainly focused on the molecular KS relation. In the rest of this letter, we   only discuss this molecular version. Measurements integrated at the full-galaxy scale up to z$\sim$2.5 showed that most galaxies lie on the local KS relation, but at higher gas surface density \citep[e.g.,][]{Tacconi2013,Freundlich2019}. Most of these high-redshift galaxies also follow a SFR--stellar mass (M$_\star$) relation, also known as the  main sequence of star-forming galaxies \citep[e.g.,][]{Noeske2007,Elbaz2007}. The normalization of this main sequence increases rapidly with increasing redshift \citep[e.g.,][]{Schreiber2015} together with the gas fraction  \citep[e.g.,][]{Magdis2012b,Saintonge2013,Bethermin2015a}, suggesting that the larger gas reservoirs are driving the higher specific SFR (sSFR=SFR/M$_\star$) observed at high z.
 
However, a small population of starbursts with a $\Sigma_{\rm SFR}$ excess compared to the KS relation of main-sequence galaxies was found by  \citet{Genzel2010} and \citet{Daddi2010b}, among others. Resolved studies of high-z starbursts confirmed the $\Sigma_{\rm SFR}$ excess at sub-galactic scale in these systems \citep[e.g.,][]{Freundlich2013,Rawle2014,Hodge2015}. Another population of starbursts with an SFR excess was also identified above the main sequence \citep[e.g.,][]{Rodighiero2011}. The starburst populations observed in both relations are suspected to be driven by mergers \citep[e.g.,][]{Sargent2014,Cibinel2019}.

To date, the KS relation in z$>$2.5 main-sequence galaxies has remained unexplored. However, the Atacama Large Millimeter Array (ALMA) has opened new perspectives to explore earlier times. In particular, the bright 158\,$\mu$m rest-fame [CII] line is now easily observable from the ground at z$\gtrsim$4, and can be used as a gas tracer \citep[e.g.,][]{Zanella2018}. Recently, \citet{Vallini2024} published a study on the KS relation in five z$\sim$7 bright Lyman beak galaxies (LBGs), but the main-sequence nature of these galaxies remains unclear. The ALMA large program to investigate [CII] at early times (ALPINE, \citealt{Le_Fevre2020,Bethermin2020,Faisst2020}) built a sample of 118 main-sequence galaxies at 4$<$z$<$6, observed in [CII] and continuum at low angular resolution ($\sim$1", marginally or not resolved). In this sample, \citet{Dessauges2020} found a flattening of the evolution of the gas fraction with redshift, similar to that observed for the  sSFR \citep{Khusanova2021}. In contrast, \citet{Jones2021} and \citet{Romano2021} estimated that a high fraction ($\sim$40\,\%) of these objects exhibits morpho-kinematical signatures of mergers despite being on the main sequence, suggesting that the mechanisms driving star formation in the z$\gtrsim$4 Universe may differ from lower redshifts.
  
In this letter we explore the KS relation at z$\sim$4.5 using a sample of four ALPINE galaxies followed up at higher resolution ($\sim0.3$\,arcsec, 2\,kpc) by ALMA. In Sect.\,\ref{sect:obs} we describe our observations and the data analysis. We then present our new results on the KS relation in Sect.\,\ref{sect:results}. Finally, we discuss them and conclude in Sect.\,\ref{sect:discussion}. We assume a flat $\Lambda$CDM cosmology (h=0.7, $\Omega_\Lambda$=0.7, $\Omega_m$=0.3) and a Chabrier initial mass function (IMF).

\begin{table*}
\centering
\caption{\label{tab:data} Summary of the observations and achieved performance. The $\sigma_{\rm cont}$ and $\sigma_{\rm [CII]}$ column are the noise of the continuum map and the [CII] moment-0 map, respectively. The 5$\sigma$ $\Sigma_{\rm gas}$ limit is derived from the  gas surface density map obtained after applying the conversion from Eq.\,\ref{eq:gas_conv}.}
\begin{tabular}{llllllll}
\hline
\hline
Source & RA & Dec & z$_{\rm [CII]}$ & beam size & $\sigma_{\rm cont}$ & $\sigma_{\rm [CII]}$ & 5$\sigma$ $\Sigma_{\rm gas}$ limit \\
& h:min:s & deg:min:s & & arcsec$^2$ & $\mu$Jy/beam & Jy\,km\,s$^{-1}$\,beam$^{-1}$ & M$_\odot$/pc$^2$\\
\hline
DC818760 & 10:01:54.86 & +2:32:31.54 & 4.5613 & 0.30$\times$0.23 & 31 & 0.068 & 2000 \\
DC873756 & 10:00:02.71 & +2:37:40.20 & 4.5457  & 0.31$\times$0.25 & 26 & 0.077 & 1940 \\
VC5101218326 & 10:01:12.50 & +2:18:52.72 & 4.5739  & 0.29$\times$0.23 & 31 & 0.074 & 1920\\
VC5110377875 & 10:01:32.33 & +2:24:30.41 & 4.5505 & 0.40$\times$0.35 & 17 & 0.037 & 420\\
\hline
\end{tabular}
\end{table*}

\section{Observations and data analysis}

\label{sect:obs}

We observed the three brightest ALPINE objects (for a full discussion of the selection, see \citealt{Devereaux2023}) in [CII] (DEIMOS\_COSMOS\_818760, DEIMOS\_COSMOS\_873756, vuds\_cosmos\_5101218326),\footnote{Hereafter, we refer to DEIMOS\_COSMOS\_ as DC and vuds\_cosmos\_ as VC.} using the C43-3 ($\sim$0.4", 73\,min) and C43-5 ($\sim$0.2", 189\,min) configurations (2019.1.00226.S). This is only a small fraction (17\,\%) of the 30\,h initially planned to study the dynamics, but it is sufficient for the goal of this letter. In addition, we observed a massive and bright ALPINE rotator candidate VC5110377875 (2022.1.01118.S) in C43-4 ($\sim$0.3") configuration during 286\,min. These observations were supposed to be followed by higher-resolution observations (C43-7, $\sim$0.07"), which were not completed.

The data were calibrated by the standard observatory pipeline, and imaged using the common astronomy software applications for radio astronomy (CASA; \citealt{CASA2022}). We produced continuum maps after excluding the [CII]-contaminated channels and the  [CII] moment-0 maps (i.e., velocity-integrated line flux maps) after subtracting the continuum in Fourier space. For DC818760, DC873756, and VC5101218326 we also included the compact-configuration visibilities from the initial ALPINE observations to recover the large-scale components. This was not necessary for VC5110377875 since the full ALPINE integrated flux is already recovered by the new high-resolution observations alone. The data reduction is described in detail in a companion paper focusing on the morpho-kinematical analysis of these objects \citep{Devereaux2023}. The source properties and the achieved ALMA performance are listed in Table\,\ref{tab:data}.

The [CII] flux is then used to derive the gas mass, while the rest-frame far-infrared continuum provides the dust obscured SFR. While initially considered to be  an SFR tracer \citep[e.g.,][]{de_Looze2014,Capak2015}, recent theoretical and observational studies pointed out that [CII] is more tightly connected to the molecular gas mass \citep[e.g.,][]{Zanella2018,Madden2020,Vizgan2022,D_Eugenio2023,Ramambason2023}, and correlates with SFR through the KS relation \citep[e.g.,][]{Ferrara2019}. The [CII] line also comes from the neutral and ionized medium, but their contribution is expected to be small in massive high-redshift galaxies \citep[e.g.,][]{Vizgan2022}. In addition, starbursting systems with short depletion timescales (M$_{\rm gas}$/SFR$\sim$100\,Myr) tend to have low [CII]-to-IR luminosity ratios \citep{Diaz-Santos2013,Gullberg2015}, which should not be the case for an SFR tracer. A theoretical discussion about this [CII] deficit in starbursts can be found in \citet{Vallini2021}. Finally, integrated [CII]-based gas masses of the ALPINE sample agree with dust-based measurements and dynamical estimates after subtracting the stellar masses \citep{Dessauges2020}. A systematic comparison of the various tracers (CO, [CI], [CII], and dust) in high-z lensed dusty star-forming galaxies, with higher SFRs than our targets by a factor of 2--50, also found a good agreement between them \citep{Gururajan2023}. Contrary to CO, the [CII] line luminosity has a weak dependence on metallicity and is sensitive to CO-dark gas as shown by studies in nearby low-metallicity dwarfs \citep{Madden2020,Ramambason2023}.

The gas surface density ($\Sigma_{\rm gas}$) is derived from the [CII] moment-0 map $m_{\rm [CII]}$ in Jy\,km\,s$^{-1}$\,beam$^{-1}$ using
\begin{equation}
\label{eq:gas_conv}
\Sigma_{\rm gas} = \alpha_{\rm [CII]} \frac{1}{D_A^2 \Omega_{\rm beam}} \left (1.04 \times 10^{-3} \frac{\rm L_\odot \, s}{\rm GHz \, Mpc^2 \, Jy \, km} \right ) D_L^2 \nu_{\rm obs}  m_{[CII]},
\end{equation}
where $\alpha_{\rm [CII]}$ is the [CII]-to-gas conversion factor (31 M$_\odot$/L$_\odot$, \citealt{Zanella2018}). This conversion factor has been cross-calibrated using CO up to z$\sim$2 to measure the molecular gas mass. If the small expected fractional contribution from other phases to the [CII] luminosity  in their sample and in our objects are similar, we should thus obtain directly a molecular gas mass corrected from the contribution of the rest of the interstellar medium. The parameter $D_A$ is the angular diameter distance and $\Omega_{\rm beam}$ the solid angle of the synthesized beam defined as the integral of a beam after normalizing its peak to unity, $D_A^2 \Omega_{\rm beam}$ being the physical area associated with the synthesized beam. The following factors in Eq.\,\ref{eq:gas_conv} correspond to the conversion from line flux to luminosity \citep{Solomon1992} with $D_L$ being the luminosity distance and $\nu_{\rm obs}$ the observed frequency.

The SFR surface density ($\Sigma_{\rm SFR}$) is the second quantity involved in the KS relation. In galaxies with a non-negligible dust content as ALPINE galaxies \citep{Fudamoto2020}, we can estimate the total SFR by combining the obscured SFR probed by the far-infrared (SFR$_{\rm IR}$) and unobscured SFR seen in the UV (SFR$_{\rm UV}$). The total SFR is the sum of these two values, and can be derived using \citep[][]{Madau2014}
\begin{equation}
\label{eq:SFRsum}
SFR = SFR_{\rm UV} + SFR_{\rm IR} = \kappa_{\rm UV} L_{\rm UV} + \kappa_{\rm IR} L_{\rm IR},
\end{equation}
where $L_{\rm UV}$ is the rest-frame 1500 $\AA$ luminosity and $L_{\rm IR}$ the total 8-1000\,$\mu$m IR luminosity; $\kappa_{\rm UV}$ and $\kappa_{\rm IR}$ are the conversion factors with a value of $1.02\times10^{-10}$\,M$_\odot$/L$_\odot$ and $1.47\times10^{-10}$\,M$_\odot$/L$_\odot$ after converting to the Chabrier IMF.

The 158\,$\mu$m rest-frame continuum map ($m_{158}$) in Jy/beam is converted into obscured SFR surface density ($\Sigma_{\rm SFR_{\rm IR}}$) using
\begin{equation}
\label{eq:sfrir_conv}
\Sigma_{\rm SFR_{\rm IR}} = \kappa_{\rm IR} \frac{1}{D_A^2 \Omega_{\rm beam}} \frac{L_{\rm IR}}{L_{158}} \nu_{\rm cont} \frac{4 \pi D_L^2}{1+z} m_{158},
\end{equation}
where $\nu_{\rm cont}$ is the rest-frame continuum frequency, and $\frac{L_{\rm IR}}{L_{158}}$ is the ratio of the total luminosity to the 158\,$\mu$m rest-frame monochromatic luminosity. We use the value of 1/0.113 computed by \citet{Bethermin2020} based on \textit{Herschel} stacking. This constant conversion factor assumes implicitly a constant dust temperature in our objects, and variations could lead to systematic effects on $\Sigma_{\rm SFR_{\rm IR}}$ estimates \citep{Cochrane2022}. Unfortunately, high-resolution observations at higher frequency to constrain the dust temperature in each line of sight are still out of reach even with ALMA.

To produce UV rest-frame maps, we started from cutouts of the HST COSMOS mosaics \citep{Koekemoer2007} in the F814W filter ($\sim$ 1480\,$\AA$ rest-frame), and converted the instrumental units into physical units (M$_\odot$\,yr$^{-1}$\,kpc$^{-2}$) using $\kappa_{\rm UV}$, the luminosity distance, and the physical area at z$\sim$4.5 corresponding to the HST pixel. To match the ALMA resolution, we then convolved the HST map by an elliptical Gaussian kernel with a major-axis width $\sigma_{\rm maj}^{\rm ker} = \sqrt{ (\sigma_{\rm maj}^{\rm ALMA})^2 - \sigma_{\rm HST}^2}$ (same formula for the minor axis), while ensuring that the normalization preserves the surface density. We tested this procedure on the HST PSF and found that the recovered beam is similar to the ALMA beam with an accuracy better than 10\,\%.

\begin{figure*}
\begin{tabular}{ccc}
\includegraphics[width=6cm]{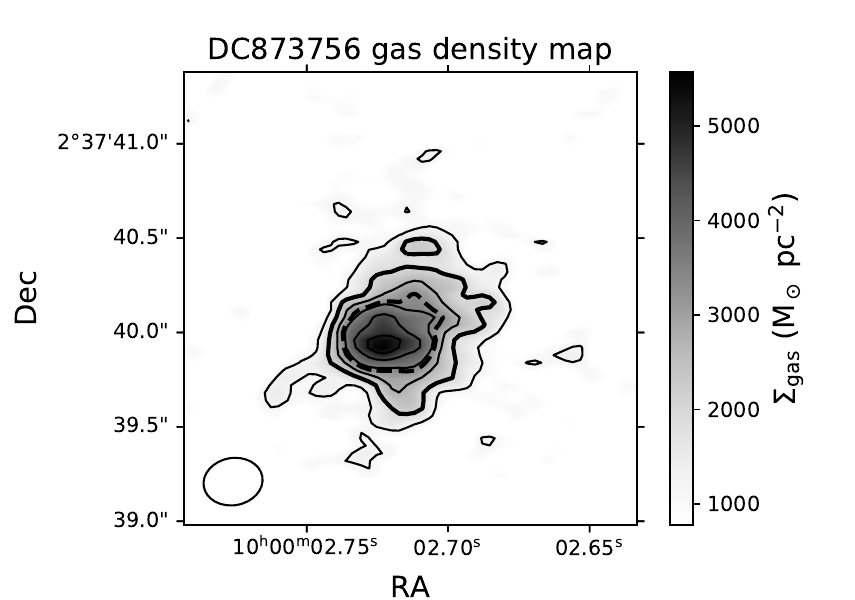} & \includegraphics[width=6cm]{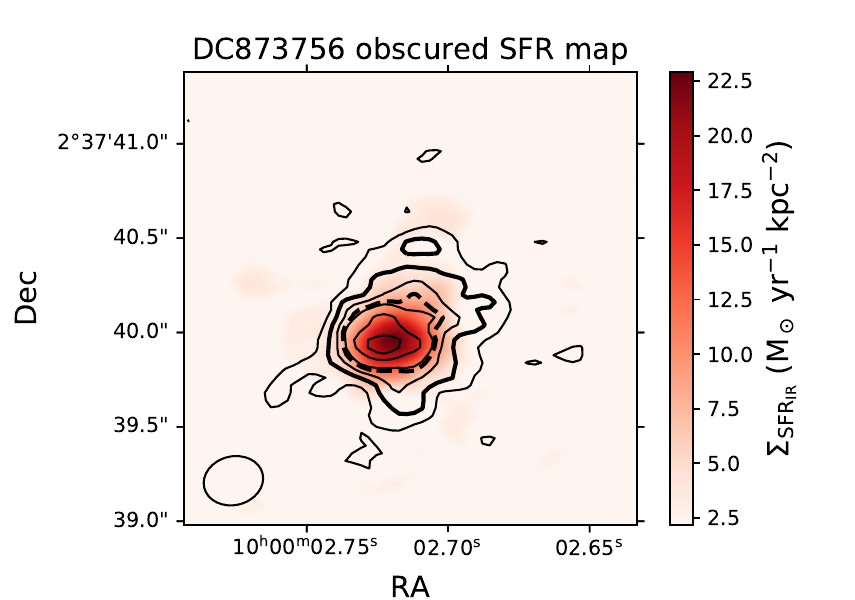} & \includegraphics[width=6cm]{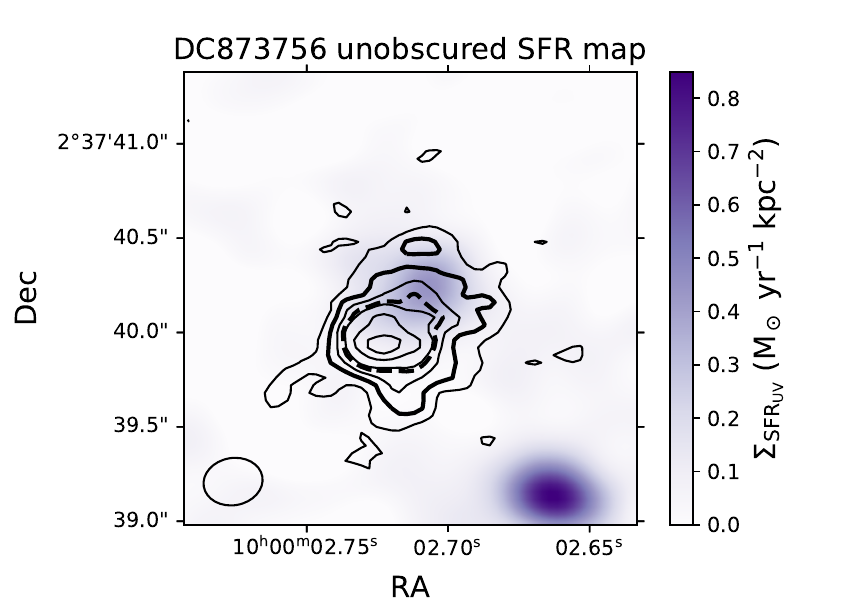} \\
\end{tabular}
\caption{\label{fig:maps_DC873756} Surface density maps of DC873756: Gas (\textit{left}, traced by [CII]), obscured  FIR SFR (\textit{center}, traced by the 158\,$\mu$m rest-frame continuum), and unobscured UV SFR (\textit{right}, traced by the F814W observer-frame continuum). The black contours are the (3+2k)\,$\sigma$ levels (k$\geq$0) of  the gas surface density. The thicker solid and dashed lines, respectively, are used  to highlight the 5$\sigma$ $\Sigma_{\rm gas}$ limit and the border between the low- and high-density regions used in our analysis (see Sect.\,\ref{sect:results}). The ALMA synthesized beam size is shown in the lower left corner. The unobscured SFR surface density map based on HST data is convolved by a Gaussian kernel to match the ALMA resolution. }
\end{figure*}

In Fig.\,\ref{fig:maps_DC873756} we present the surface density maps of DC873756 as an example, while the other sources are shown and briefly discussed in Appendix\,\ref{sect:other_maps}. We note that the gas map has a very high S/N, and only the galaxy core is detected in the obscured SFR map. We also note that the obscured and unobscured SFR surface density maps exhibit different morphologies with the UV star formation coming mainly from the diffuse gas extension in the northwest and not the IR-bright core. The southwest bright UV blob has a robust photo-z of 3.4$\pm0.1$ \citep{Weaver2022}, and is thus probably not related to our target. This highlights the necessity to have access to both the obscured and unobscured star formation in our analysis. This source is thus asymmetric with a dusty star-forming core slightly offset in the southeast direction and a diffuse and less obscured extension in the northwest, illustrating the complexity of the morphologies at this redshift (see \citealt{Devereaux2023} for more details).

\begin{figure}
\begin{tabular}{cc}
\includegraphics[width=4.3cm]{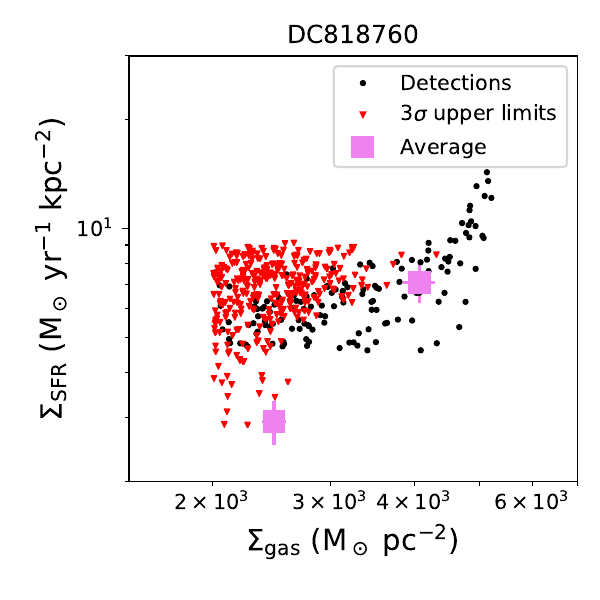} & \includegraphics[width=4.3cm]{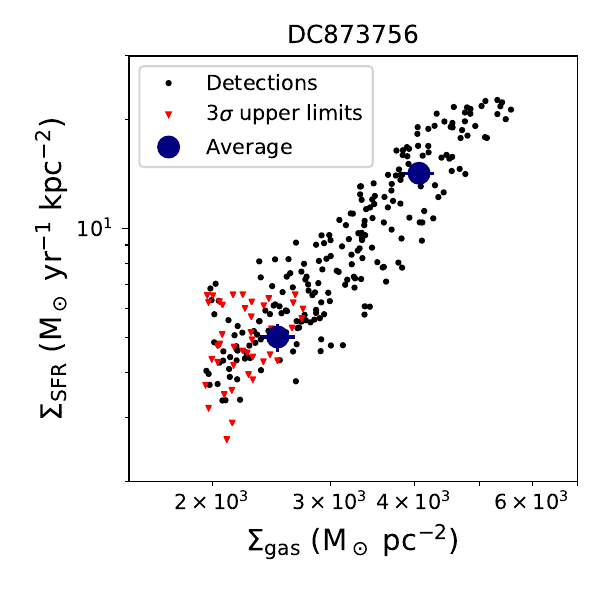}\\
\includegraphics[width=4.3cm]{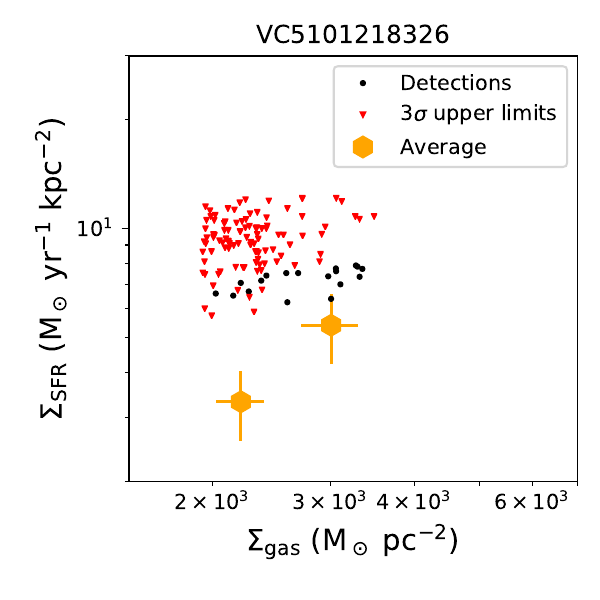} & \includegraphics[width=4.3cm]{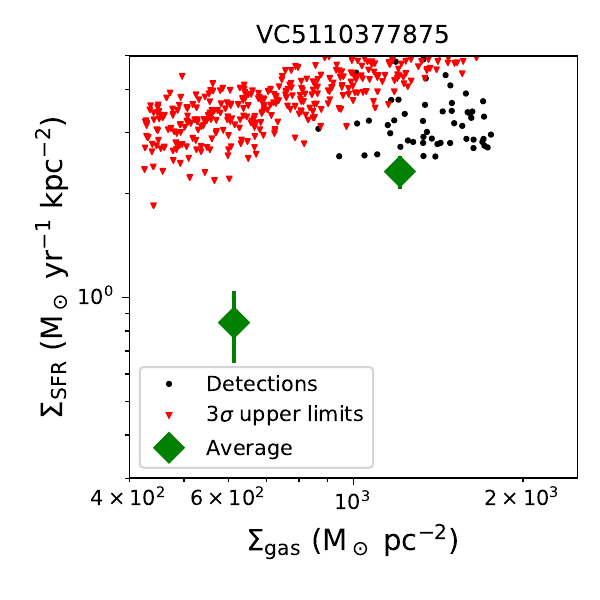}\\
\end{tabular}
\caption{\label{fig:pixbypix} Pixel-by-pixel KS relation between the SFR and gas surface density. Each panel corresponds to a source (\textit{upper left}: DC818760, \textit{upper right:} DC873756, \textit{lower left}: VC5101218326, \textit{lower right:} VC5110377875). The small black filled circles correspond to lines of sight with a $>$3\,$\sigma$ signal on the SFR surface density, and the small red downward triangles to 3\,$\sigma$ upper limits. The large filled symbols (same as in Fig.\,\ref{fig:KS}) correspond to the average value for all the pixels in a given gas surface density bin (see Sect.\,\ref{sect:results} and Table\,\ref{tab:results}). }
\end{figure}

\section{Results}

\label{sect:results}

We first investigated the KS relation on a pixel-per-pixel basis (0.06" in VC5110377875 and 0.04" in the other galaxies). Since our $\Sigma_{\rm gas}$ map is by far the deepest, we restricted our study to the regions were the gas is detected in [CII] at better than 5\,$\sigma$ (see Table\,\ref{tab:data}) to robustly avoid working on noise spikes in the source outskirts. We then summed the $\Sigma_{\rm SFR_{\rm IR}}$ and $\Sigma_{\rm SFR_{\rm UV}}$ maps to obtain the $\Sigma_{\rm SFR}$ maps, and assumed a quadratic combination of the noise. In our objects, the obscured SFR contributes 75-99\,\% of the total SFR. In Fig.\,\ref{fig:pixbypix} we show our results for each source. While in DC873756 most of the lines of sight have a $\Sigma_{\rm SFR}$ signal above 3\,$\sigma$, this is not the case in the three other sources. The risk of bias is very high in the $\Sigma_{\rm gas}$ regime where a small fraction of pixels are detected in $\Sigma_{\rm SFR}$. In addition, in interferometry, the data points from the neighboring pixels are not independent because the noise is correlated at the scale of the synthesized beam. In this letter we thus only focus on deriving unbiased average quantities, which unfortunately prevents us from studying the scatter.

For each source, we defined two different regions based on their $\Sigma_{\rm gas}$. We cut the range between the 5$\sigma$ limit and the  maximum in two bins of the same logarithmic size, building two different regions with a lower and a higher $\Sigma_{\rm gas}$. The high-density region tends to be in the center, while the lower-density region forms a ring around it (see Fig.\,\ref{fig:maps_DC873756}). However, the disturbed morphologies of our systems lead to rather complex shapes, which justifies a posteriori not   using radial profiles in this  object type. Finally, we computed the mean $\Sigma_{\rm gas}$ and $\Sigma_{\rm SFR}$ in each region of each object. In Appendix\,\ref{sect:simus} we describe a simulation validating our method. This simulation shows that we can recover without bias the intrinsic relation in the presence of noise. However, this simulation does not include the potential complex systematic effects from spatial variations of the conversion factors used in Eqs.\,\ref{eq:gas_conv}, \ref{eq:SFRsum}, and \ref{eq:sfrir_conv}.

To derive uncertainties on these quantities, we moved randomly the two regions in a noisy area of the map using the same offset for both regions, and measured our observables using the same method as previously. This process was repeated 10\,000 times, and the standard deviation of the results provides the uncertainty. The results are summarized in Table\,\ref{tab:results}. We also computed the typical correlation between the noise realizations in the two regions and found a Pearson correlation coefficient of 0.5$\pm$0.1. This is expected since the two regions have a common border, and the interferometric noise is correlated at the scale of the synthesized beam. In Fig.\,\ref{fig:pixbypix} these average values are located at the middle of the cloud of points when most of the pixels are detected (DC873756 and high-density bin of DC818760), but lie significantly below otherwise. This is not surprising since in the regime of low average S/N the detections are biased  toward positive outliers of the KS relation and the instrumental noise.

\begin{table}
\caption{\label{tab:results} Average gas ($\langle \Sigma_{\rm gas} \rangle$) and SFR ($\langle \Sigma_{\rm SFR} \rangle$) surface density measured in two $\Sigma_{\rm gas}$ bins for our four objects. The uncertainties are derived using the  Monte Carlo simulations described in Sect.\,\ref{sect:results}. The uncertainties are based only on the instrumental noise.}
\begin{tabular}{llll}
\hline
\hline
Name & $\Sigma_{\rm gas}$  range & $\langle \Sigma_{\rm gas} \rangle$ & $\langle \Sigma_{\rm SFR} \rangle$ \\
& M$_\odot$/pc$^2$ & M$_\odot$/pc$^2$ & M$_\odot$\,yr$^{-1}$\,kpc$^{-2}$ \\
\hline
DC818760 & 2005-3232 & 2468$\pm$101 & 2.9$\pm$0.4\\
DC818760 & 3232-5208 & 4068$\pm$219 & 7.1$\pm$0.9\\
DC873756 & 1945-3291 & 2500$\pm$151 & 5.0$\pm$0.4\\
DC873756 & 3291-5570 & 4062$\pm$215 & 14.2$\pm$0.6\\
VC5101218326 & 1926-2590 & 2203$\pm$182 & 3.3$\pm$0.7\\
VC5101218326 & 2590-3481 & 3003$\pm$295 & 5.4$\pm$1.2\\
VC5110377875 & 424-862 & 613$\pm$37 & 0.8$\pm$0.2\\
VC5110377875 & 862-1755 & 1210$\pm$50 & 2.3$\pm$0.3\\
\hline
\end{tabular}
\end{table}

\begin{figure}
\includegraphics[width=8.5cm]{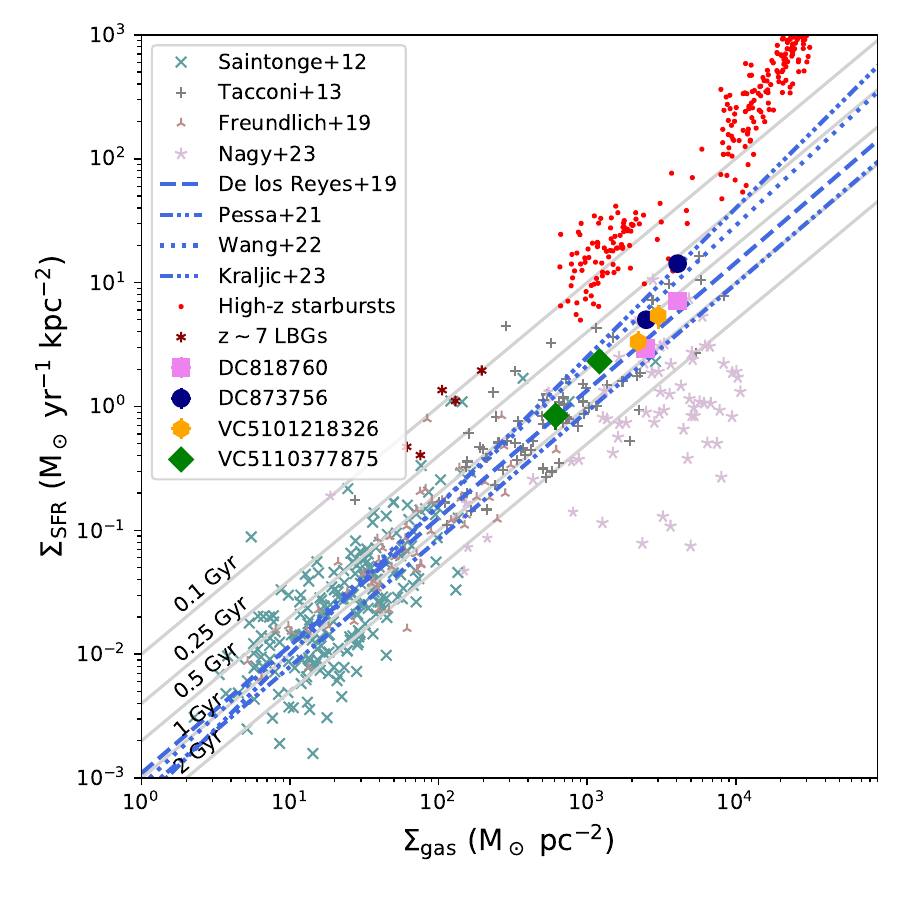}
\caption{\label{fig:KS} Kennicutt-Schmidt relation between the SFR and gas surface density. The averaged data points from DC818760, DC873756, VC5101218326,  and VC5110377875 are shown as violet squares, blue filled circles, yellow hexagons, and green diamonds, respectively. Our results are compared with the CO measurements of the global KS relation from the low-z COLD GASS sample (\citealt{Saintonge2012}, crosses) and at redshifts up to z$\sim$2.5 from the PHIBBS \citep[][plus signs]{Tacconi2013} and PHIBBS2 \citep[][three-branch stars]{Freundlich2019} programs, together with the resolved KS relation in z$\sim$1 lensed main-sequence galaxies \citep[five-branch stars]{Nagy2023}. Also shown are the global (\citealt{De_los_Reyes2019}, dashed line) and resolved (\citealt{Pessa2021}, two-dot-dashed line) KS relations measured in the local Universe, the high-redshift global KS relation obtained by ALMA CO stacking (\citealt{Wang2022}, dotted line), and the global KS relation from simulations at z$\sim$4 (\citealt{Kraljic2023}, three-dot-dashed line). The thin gray lines indicate the various depletion timescales. The red points are resolved measurements in high-z starbursts by \citet{Hodge2015}. The brown six-branch stars are the measurements in z$\sim$7 LBGs by \citet{Vallini2024}. The error bars include only the uncertainties from the instrumental noise, but not from the calibration ($\sim$10\,\% for ALMA) or from the conversion factors used in Sect.\,\ref{sect:obs}. They are often smaller than the symbols.}
\end{figure}

In Fig.\,\ref{fig:KS} we present a synthesis of our results together with a compilation of previous works. All our data points but one are consistent with a 0.5-1\,Gyr depletion timescale expected for massive main-sequence galaxies at z$\sim$4.5 \citep{Scoville2017,Dessauges2020}. The only outlier (285$\pm$20\,Myr) is the central high-density region of DC873756, which has the highest obscured SFR fraction (SFR$_{\rm IR}$/SFR$_{\rm tot}$ = 98.7$\pm$0.3\,\%) of the sample, and could thus hide a mild but heavily obscured starburst.

We compared our results with CO measurements of the global KS relation from the nearby COLD GASS sample \citep{Saintonge2012} and the z$\lesssim$2.5 PHIBBS and PHIBBS2 samples \citep{Tacconi2013,Freundlich2019}. We find that our objects are located in the scatter of these previous studies ($\sim$0.3\,dex corresponding to a factor of 2), but with a higher gas surface density than most of the nearby sample by almost two orders of magnitude. Similarly, our sources are within a factor of 2 of the global KS relation fits by \citet{De_los_Reyes2019} at low redshift and by \citet{Wang2022} at high redshift, and the resolved low-z KS fit by \citet{Pessa2021}. In contrast, most of the measurements of \citet{Nagy2023} at the scale of 1.6\,kpc (compared to $\sim$2\,kpc in our analysis) in two strongly lensed z$\sim$1 galaxies have significantly lower $\Sigma_{\rm SFR}$ at a given $\Sigma_{\rm gas}$ than other analyses; this is discussed in their paper without converging on a final explanation. Since their Cosmic Snake measurements agree with those of  \citet{Pessa2021} and with our results, but not their measurements in the A521 galaxy, it could suggest that the the A521 galaxy  is an outlier. In contrast, our sample of main-sequence galaxies have significantly lower $\Sigma_{\rm SFR}$ at a given $\Sigma_{\rm gas}$ than typical high-redshift starbursts, which have an average depletion timescale of 100\,Myr \citep{Sharon2013,Rawle2014,Hodge2015}. The z$\sim$7 LBGs studied by \citet{Vallini2024} exhibit a similar behavior, suggesting that they could also be starbursting systems. Finally, we compared our results with the KS relation of \citet[][fit for z=4 and $10^8 <M_\star / M_\odot < 10^{9.5}$]{Kraljic2023} from the hydrodynamical simulation NewHorizon \citep{Dubois2021}. Their results suggest a slightly higher normalization (by a factor of $\sim$1.5) than our measurements. However, objects as massive as our targets are not found in the limited volume of their simulated box, and low- and high-mass high-z galaxies could exhibit different behaviors.

\section{Discussion and conclusion}

\label{sect:discussion}

Our study suggests that main-sequence galaxies follow the same KS relation from z=0 to z=4.5, but exhibit a strong increase in their gas density with increasing redshift driving a rise in their SFR density. This universality of the KS relation for main-sequence galaxies might seem surprising considering how different local and z$>$4 galaxies are, for example in term of gas fraction or sSFR. Since a significant fraction of our sample \citep{Devereaux2023}, and more globally of z$\sim$5 massive main-sequence galaxies \citep[e.g.,][]{Romano2021}, are identified as possible mergers by morpho-kinematical analyses, this opens the question of a possible decoupling of the merging and starburst events in high-redshift galaxies, but also of the timescale of the SFR enhancement and the associated merger stage. Theoretical works suggest that high-z mergers could often be  inefficient in increasing star formation \citep[e.g.,][]{Fensch2017} because their turbulence-dominated medium is incapable of producing high-density regions, decreasing dramatically the gas depletion timescale \citep{Segovia_Otero2022}. However, since extremely star-forming major starbursts are routinely observed in the submillimeter domain \citep[e.g.,][for a review]{Hodge2020}, we need to better understand the mechanism at play in the high-redshift Universe, and how they differ or not from low-z galaxies. Submillimeter galaxy selections could be either particularly efficient at finding the hypothetically rare and short starbursts at this epoch, or these more extreme objects may just have different physical conditions than our targets (e.g., different physical or dynamical properties of the pre-merger components). Objects, such as  DC873756 that has  a shorter depletion timescale in its obscured core, are particularly interesting for studying the intermediate regime.

Our analysis is a very first step. We need to confirm these results on larger samples and with other SFR (e.g., H$\alpha$ with JWST) and gas tracers, since [CII] has not been tested at sub-galactic scale in the high-redshift Universe yet. Ideally, we should also push toward lower masses with potentially less mature systems and lower metallicities. Deeper ALMA observations would also allow us to detect the dust continuum in most lines of sight providing a local estimate of $\Sigma_{\rm SFR_{\rm IR}}$, and measure the scatter on the KS relation. Finally, higher-resolution observations will test whether the KS relation breaks at smaller scales in main-sequence galaxies, as discussed in \citet{Nagy2023} at z$\sim$1.

\begin{acknowledgements}
We thank Jackie Hodge for sharing her data compilation. This paper makes use of the following ALMA data: ADS/JAO.ALMA\#2017.1.00428L, ADS/JAO.ALMA\#2019.1.00226.S, ADS/JAO.ALMA\#2022.1.01118.S. ALMA is a partnership of ESO (representing its member states), NSF (USA) and NINS (Japan), together with NRC (Canada), MOST and ASIAA (Taiwan), and KASI (Republic of Korea), in cooperation with the Republic of Chile. The Joint ALMA Observatory is operated by ESO, AUI/NRAO and NAOJ. This work was supported by the Programme National Cosmology et Galaxies (PNCG) of CNRS/INSU with INP and IN2P3, co-funded by CEA and CNES. E.I. acknowledge funding by ANID FONDECYT Regular 1221846. The Flatiron Institute is supported by the Simons Foundation. GG acknowledges support  from the grants PRIN MIUR 2017 - 20173ML3WW\_001, ASI n.I/023/12/0 and INAF-PRIN 1.05.01.85.08. M\'ed\'eric Boquien gratefully acknowledges support from the ANID BASAL project FB210003 and from the FONDECYT regular grant 1211000. GEM acknowledges the Villum Fonden research grant 13160 “Gas to stars, stars to dust: tracing star formation across cosmic time,” grant 37440, “The Hidden Cosmos,” and the Cosmic Dawn Center of Excellence funded by the Danish National Research Foundation under the grant No. 140. This work was supported by NAOJ ALMA Scientific Research Grant Code 2021-19A (HSBA). M.R. acknowledges support from the Narodowe Centrum Nauki (UMO-2020/38/E/ST9/00077) and support from the Foundation for Polish Science (FNP) under the program START 063.2023. YF acknowledges support from JSPS KAKENHI Grant Number JP23K13149.
\end{acknowledgements}

\newpage

\bibliographystyle{aa}

\bibliography{biblio}

\begin{appendix}

\section{Gas and SFR maps of the other sources}

\label{sect:other_maps}

\begin{figure}
\includegraphics[width=9cm]{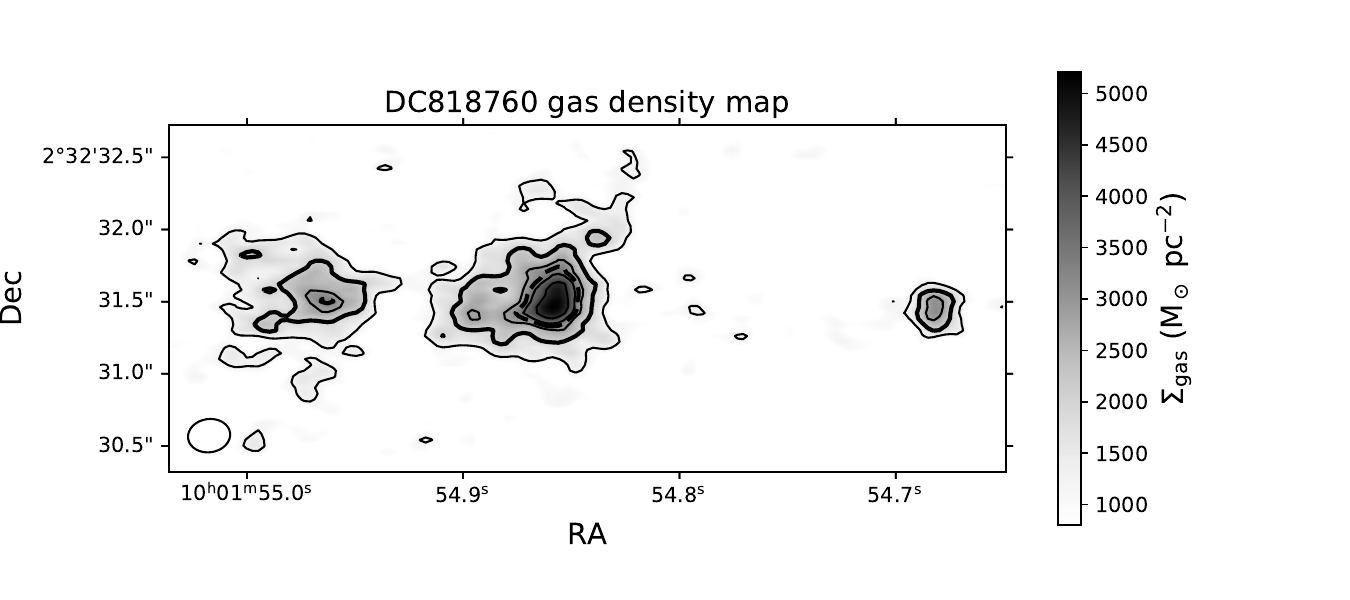} \\
\includegraphics[width=9cm]{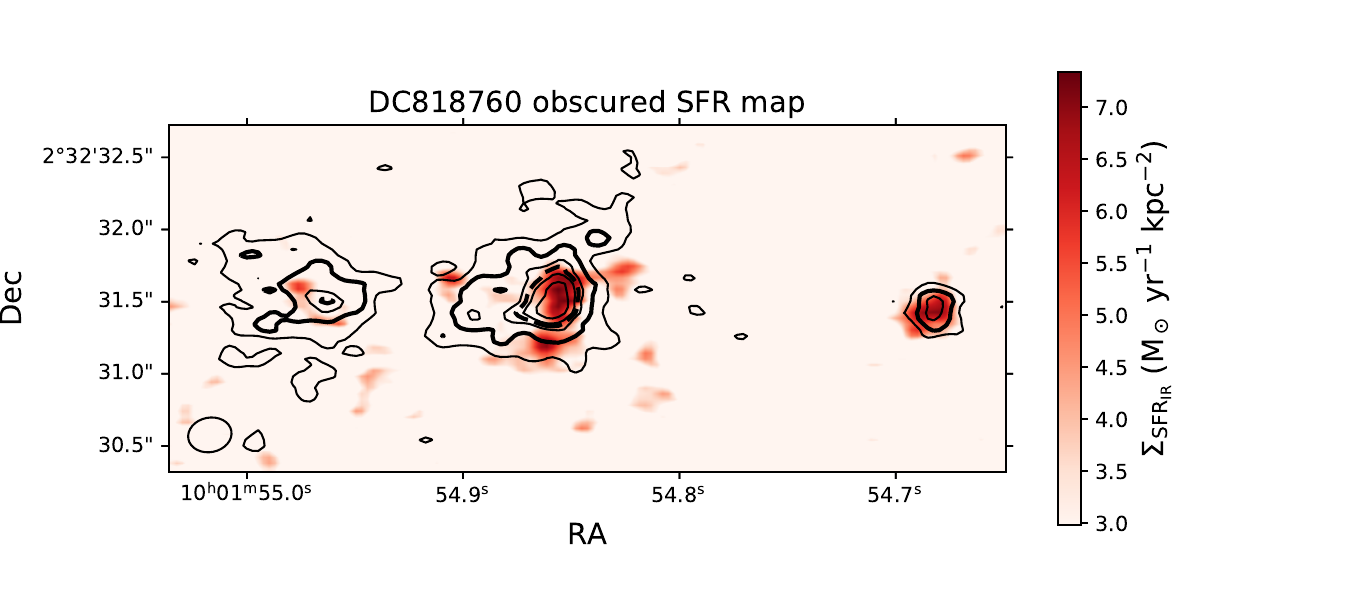} \\
\includegraphics[width=9cm]{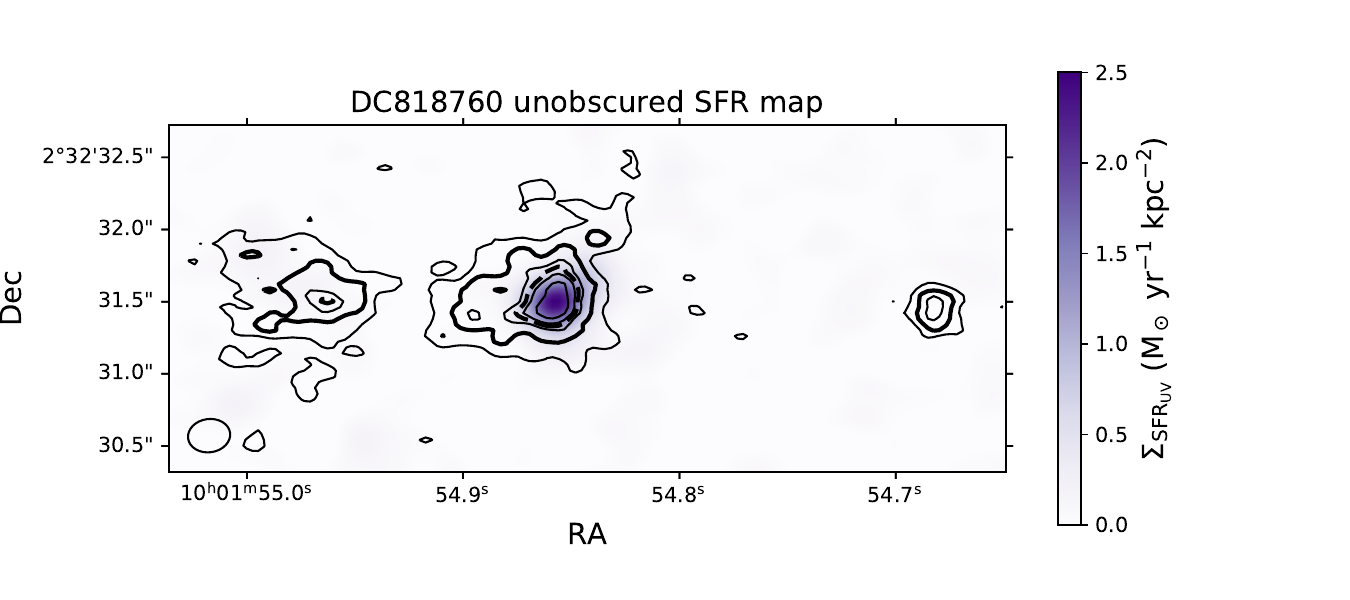}
\caption{\label{fig:maps_DC818760} Same as Fig.\,\ref{fig:maps_DC873756}, but  for DC818760.}
\end{figure}

\begin{figure}
\begin{tabular}{cc}
\includegraphics[width=4.5cm]{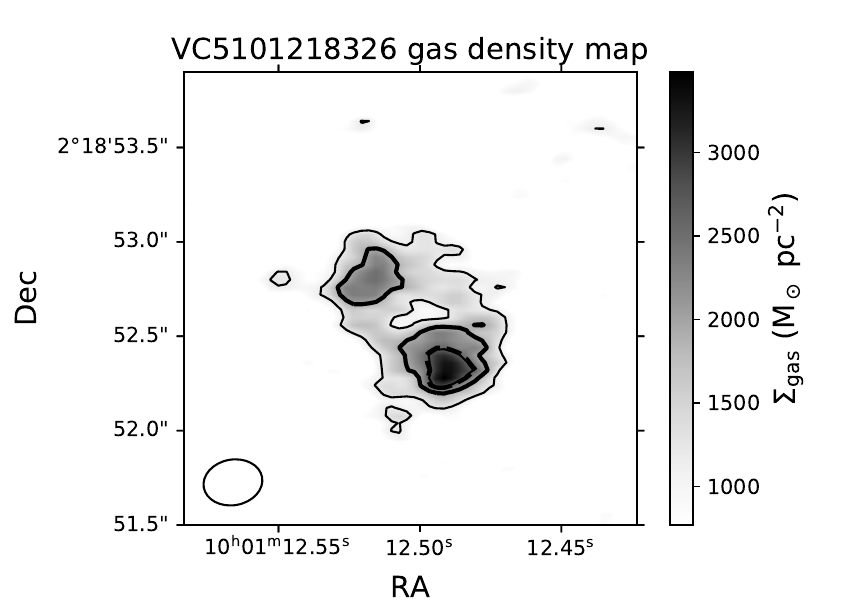} & \includegraphics[width=4.5cm]{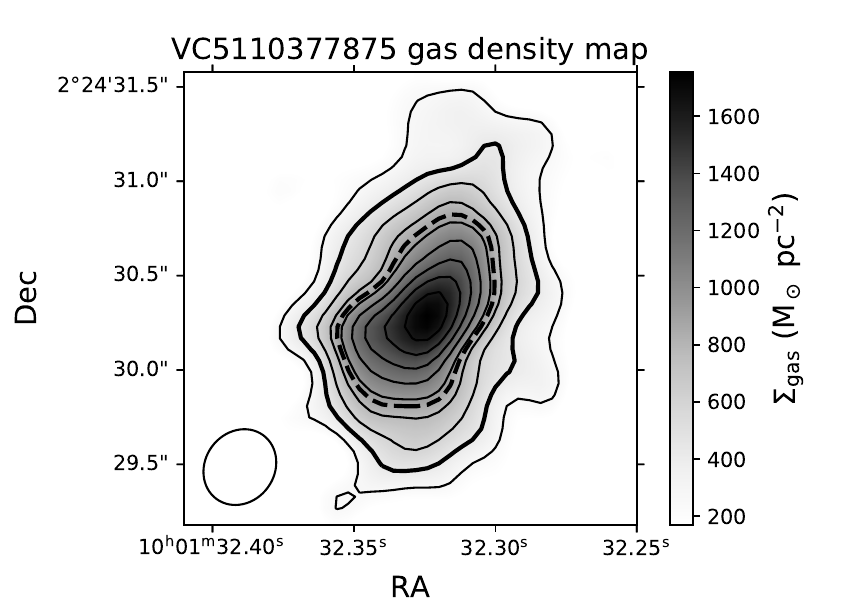} \\
\includegraphics[width=4.5cm]{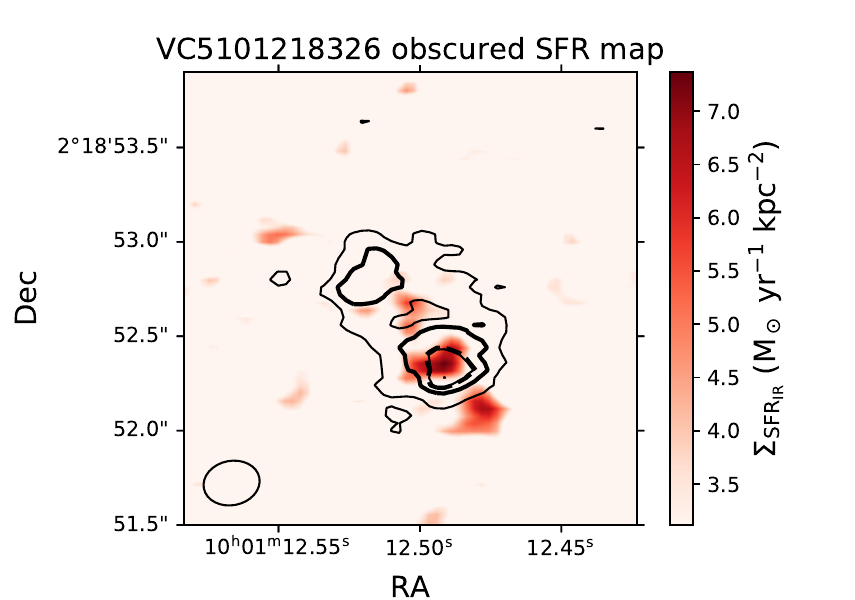} & \includegraphics[width=4.5cm]{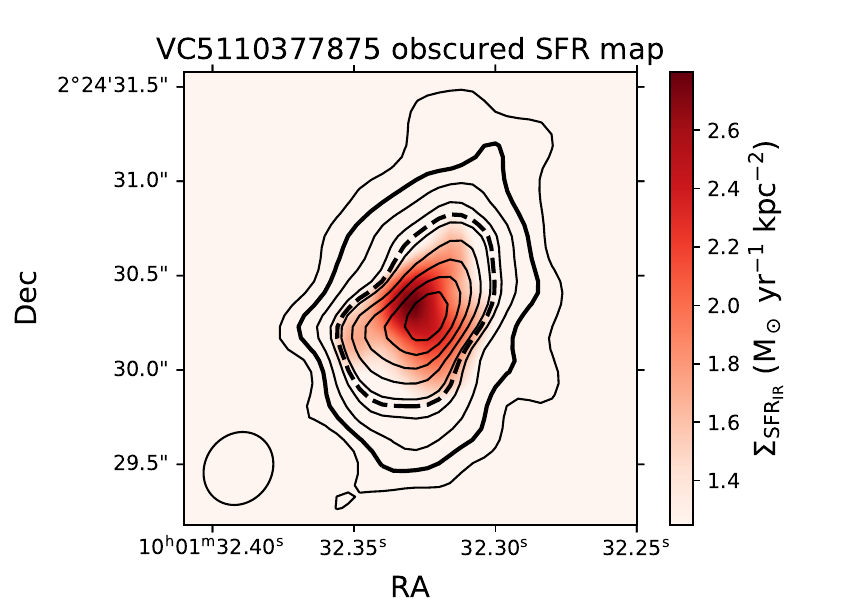} \\
\includegraphics[width=4.5cm]{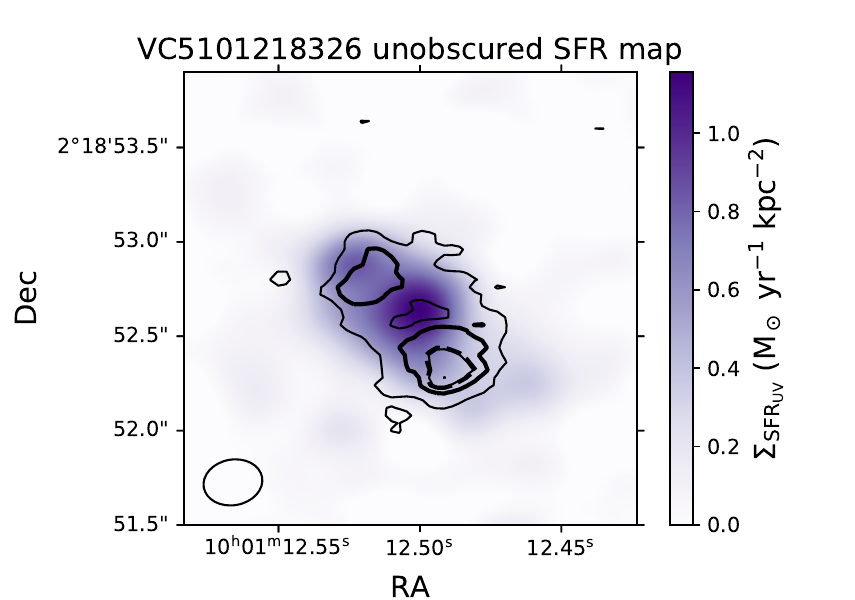} & \includegraphics[width=4.5cm]{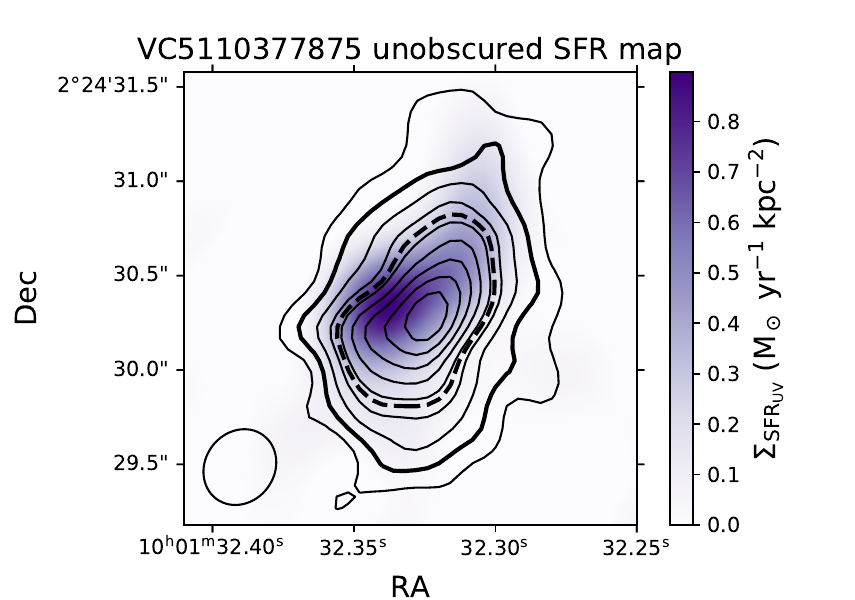} \\
\end{tabular}
\caption{\label{fig:maps_VCs} Same as Fig.\,\ref{fig:maps_DC873756}, but  for VC5101218326 (\textit{left}) and VC5110377875 (\textit{right}).}
\end{figure}

Figures\,\ref{fig:maps_DC818760} and \ref{fig:maps_VCs} show the gas and SFR density maps of DC818760, VC5101218326, and VC5110377875, which are not discussed in the main text. Similarly to DC873756, we note  that   DC818760 and VC5101218326 have very different obscured and unobscured SFR density maps, confirming the need to have access to both. For instance, both the central and western components of DC818760 host an important obscured star formation activity, but only the central one is detected in the unobscured SFR density map. The morphology and the kinematics of these two objects is discussed in a companion paper (\citealt{Devereaux2023}; see also \citealt{Jones2020} about the initial ALPINE data of DC818760). They are both identified as   solid merger candidates.   VC5110377875  will be studied  in greater detail  when the high-resolution data   reaches a sufficient depth. So far it looks like a well-behaved rotator. However, we  note a small displacement of the unobscured SFR toward the east compared to the gas.

\section{Depletion timescales measured inside apertures with various radii}

\label{sect:radial_tdep}

Since our targets have complex morphologies, we decided to measure the average $\Sigma_{\rm SFR}$ in regions selected to be in a specific range of $\Sigma_{\rm gas}$. This type of measurement is easy to reproduce in simulations. This method is particularly well suited for z$>$4 observations since the [CII] line used as a gas tracer  usually has a better S/N (by a factor of $\sim$3) than the dust continuum tracing the obscured SFR. It is not the case at lower redshift, where the gas tracers are often shallower than SFR estimators. Usually, the gas and SFR surface densities are measured in an aperture corresponding to the source size. This leads to multiple questions about the choice of wavelength or line to measure the source size, but also the exact definition of the source radius.

\begin{figure}
\includegraphics[width=8.5cm]{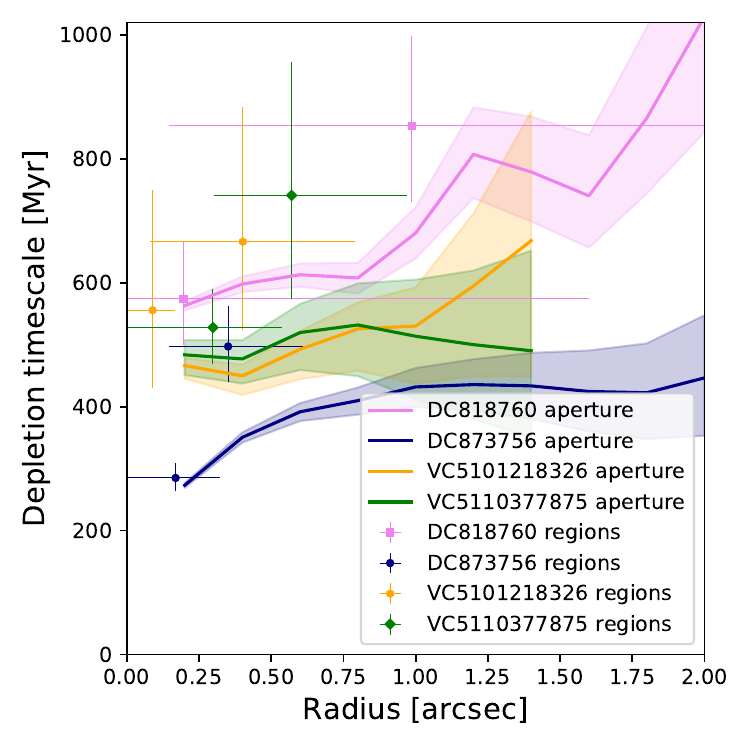}
\caption{\label{fig:radial_tdep} Gas depletion timescale measured in an aperture centered on the maximum of the [CII] emission as a function of its radius (solid lines, see Appendix\,\ref{sect:radial_tdep}). The colored areas correspond to the 1$\sigma$ uncertainties determined using the Monte Carlo sampling described in Sect.\,\ref{sect:results}. For comparison, we represent the depletion timescales measured in our two $\Sigma_{\rm gas}$ regions using the same symbols as in Fig.\,\ref{fig:KS}. The horizontal error bars correspond to the full range of angular radial distances to the [CII] maximum in the region. The high-density region systematically corresponds to the smaller radius.}
\end{figure}

To compare the two approaches, we measured the gas depletion timescale in apertures centered on the maximum of the [CII] emission with various radii. This choice may appear arbitrary, but the UV center is often not representative of the mass distribution of the system because of the dust attenuation, and to date we have no access  to high-resolution rest-frame near-infrared data probing the old stars. The uncertainties are determined using the same Monte Carlo resampling method that we used for the $\Sigma_{\rm gas}$ regions in Sect.\,\ref{sect:results}. The comparison of the two approaches is shown in Fig.\,\ref{fig:radial_tdep} (solid lines). We note that two sources have a significantly longer gas depletion timescale for large aperture radii (DC818760, DC873756), while the two others do not exhibit any significant trend (VC5101218326, VC5110377875). DC818760 is a multi-component system and the large radii include  the two neighbors, while the small radii only include the central component. It is not surprising to find slightly different timescales in the various components. As discussed in Sect.\,\ref{sect:results}, DC873756 probably hosts an obscured starbursting core with a short depletion timescale, while it is longer in the outskirts.

In Fig.\,\ref{fig:radial_tdep} we also show the gas depletion timescales found in our low- and high-$\Sigma_{\rm gas}$ regions. The x-axis position of these datapoints corresponds to the average angular radial distance to the maximum of the [CII] emission. The error bars represent the full range of distances. Since the morphologies are not centro-symmetrical, there is an overlap between the two regions in terms of radial distance. The high-density regions correspond to smaller radii and agree with aperture measurements at the 1$\sigma$ level. This is not the case of the low-density regions at larger radii for which the disagreement can reach 2\,$\sigma$. However, the aperture measurements are cumulative and includes all the pixels from the center to the edge of the aperture. The two approaches are thus not measuring exactly the same thing, and a disagreement is expected if there are longer depletion timescales in the outskirts. However, the maximum disagreement caused by the choice of definition does not exceed 50\,\%, and is thus smaller than the scatter on the relations (typically a factor of 2). Consequently, the choice of the approach used to determine surface densities does not change  the results qualitatively.

\section{Testing the method measuring the mean gas and SFR surface densities}

\label{sect:simus}

We built a simulation to test the accuracy of our method to derive mean $\Sigma_{\rm gas}$ and $\Sigma_{\rm SFR}$ (see Sect.\,\ref{sect:results}). We produced   cases based on DC873756 (hereafter case 1), for which most of the pixels in both the low and high gas density regions are detected in the combined $\Sigma_{\rm SFR}$ map, and VC5110377875 (case 2), for which most of the pixels in the low-density region are undetected. The first source is modeled by a symmetrical Gaussian with a FWHM of 0.5", and the second by a 1.5"$\times$0.5" elliptical Gaussian. We convolved these models by the associated synthesized beams in the real observations and rescaled the peak of the convolved maps to match the observed values. We thus obtained three noiseless maps ($\Sigma_{\rm gas}$, $\Sigma_{\rm SFR_{IR}}$, $\Sigma_{\rm SFR_{UV}}$) for each source. Since we applied a simple rescaling to our models, the local depletion timescale is thus constant across the simulated objects by construction. Finally, we added interferometric noise to the $\Sigma_{\rm gas}$ and $\Sigma_{\rm SFR_{IR}}$ maps after rescaling it to the observed values, and  white noise smoothed by our convolution kernel to the $\Sigma_{\rm SFR_{UV}}$ map after matching its level to the observed value.

\begin{figure}
\includegraphics[width=9cm]{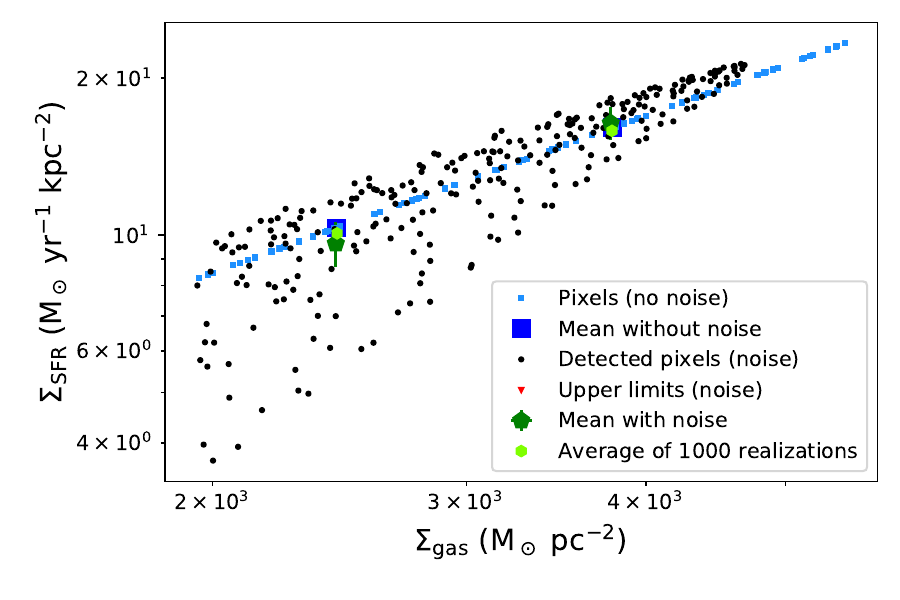}
\includegraphics[width=9cm]{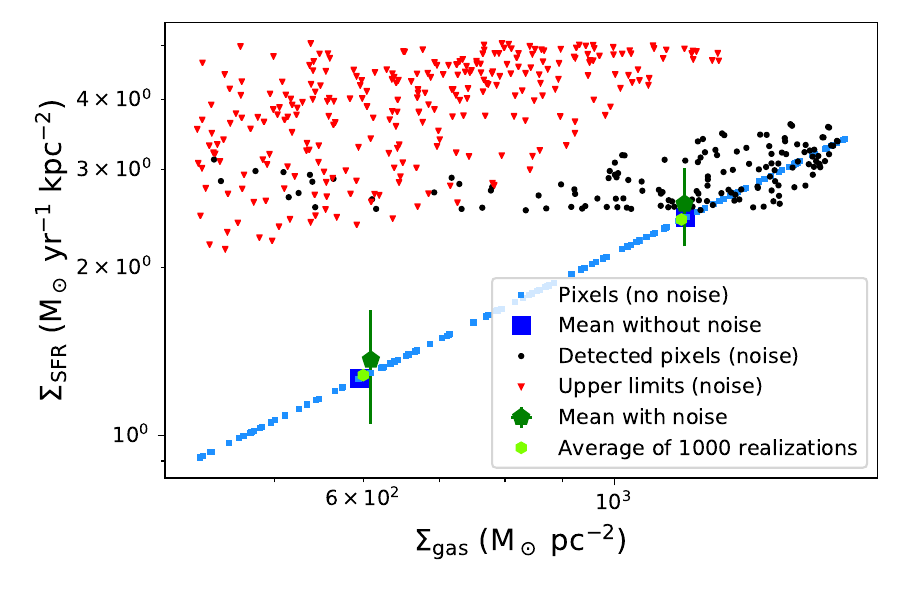}
\caption{\label{fig:simus} Kennicutt-Schmidt relation between the SFR and gas surface density in our simulation described in Appendix\,\ref{sect:simus}. The  panels correspond to two different simulated objects based on DC873756 (\textit{upper panel}) and VC5110377875 (\textit{lower panel}), illustrating different S/N regimes. The small blue squares show the pixels of the noiseless map, and the large dark blue squares are the mean computed in the low and high gas density regions (see Sect.\,\ref{sect:results}). The small black filled circles are the 5$\sigma$ SFR detections in the noisy maps, and the small red downward-facing triangles are the 3$\sigma$ upper limits of non-detections. The dark green filled pentagons are the mean values measures in the two gas density regions. The small light green hexagons are the averages of these mean measurements over 1000 noise realizations.}
\end{figure}

We then applied the same measurement process as described in Sect.\,\ref{sect:results} on both noiseless and noisy maps. The results are presented in Fig.\,\ref{fig:simus}. The KS relation in the noiseless case is our reference (light blue small squares). Since there is no noise associated with these maps, we used the noise level of the $\Sigma_{\rm gas}$ noisy map to define the region above the 5$\sigma$ density limit in which we analyzed individual pixels. The low and high gas density regions are defined in a similar way, and we derived the mean $\Sigma_{\rm gas}$ and $\Sigma_{\rm SFR}$ in each of them (dark blue large squares). As expected, there is a near perfect agreement between the individual pixels and the mean measurements in the two simulated objects.

We also analyzed the noisy maps. The various regions used in our analysis can slightly change compared to the noiseless case, since some pixels can pass above or below a threshold because of the noise. The individual detected pixels (small black filled circles) agree with the noiseless relation in case 1 where the S/N is high. In case 2 the detections agree only at the high-$\Sigma_{\rm gas}$ end, but are systematically higher than the noiseless case at the low-$\Sigma_{\rm gas}$ end. Since only pixels with $\Sigma_{\rm SFR}$ above 2.5\,M$_\odot$\,yr$^{-1}$\,kpc$^{-2}$ are detected, only the positive noise outliers are detected in the low-$\Sigma_{\rm gas}$ regime for which the average $\Sigma_{\rm SFR}$ is below this limit. In contrast, the mean values measured in the low and high gas density regions (dark green pentagons) agree at 1$\sigma$ with the noiseless relation. To check for  a possible bias, we computed their average positions over 1000 noise realizations (light green hexagons), and they agree at better than 3\,\% with the noiseless case.

Our simulation confirms that mean measurements over several well-selected regions are more effective at recovering mean $\Sigma_{\rm SFR}$ than individual pixel measurements, which are strongly biased in the low S/N regime. It is not surprising, since these regions encompass several synthesized beams and averaging them improves the S/N by approximately a factor of $\sqrt{N_{\rm sb}}$, where $N_{\rm sb}$ is the number of synthesized beams in the region.\footnote{This is   an approximation since the shape of the regions and the spatially correlated interferometric noise are both complex, but it works with a 20\,\% accuracy in our simulations.} However, this method works only because we benefit from deep  $\Sigma_{\rm gas}$ maps to select these regions.

\end{appendix}

\end{document}